\documentclass{elsart} 
\input epsf

\newcommand{\vk}{\vec{k}}
\newcommand{\vq}{\vec{q}} 
\newcommand{\vp}{\vec{p}} 
\newcommand{\vpp}{\vec{p}\,\,'\!\!\!} 

\newcommand{\vK}{\vec{K}} 
\newcommand{\vx}{\vec{x}} 
 
\newcommand{\cj}{C(j)} 
\newcommand{\ci}{C(i)} 
 
\newcommand{\tpi}{2\pi} 
\newcommand{\Mn}{M_{_N}} 
\newcommand{\ubar}{\overline{u}} 
\newcommand{\Ubar}{\overline{U}} 
\newcommand{\lam}{\lambda} 
\newcommand{\vlam}{\vec{\lambda}} 
\newcommand{\Mc}{M_{_C}} 
\newcommand{\Ma}{M_{_A}}

\newcommand{\braj}{\{j\}} 
 
\begin{document} 
\begin{center}
\title{Pauli blocking and final-state interaction in electron-nucleus  
quasielastic scattering } 
 
Lon-chang (L.C.) Liu$^{\dagger}$ 
    
{Theoretical Division, Group T-16, Mail Stop B243}\\     
{Los Alamos National Laboratory, Los Alamos, NM 87545 USA}

%\date{\today} 
%\date{December 14, 2008} 
%\maketitle 
\end{center} 

\vspace{1cm}  
\begin{center} 
{\bf Abstract}
\end{center}   
The nucleon final-state interaction in inclusive  
electron-nucleus quasielastic scattering is 
studied.   
Based on the unitarity equation satisfied by the 
scattering-wave operators,     
a doorway model is developed to take into account
the final-state interaction including 
the Pauli blocking of nucleon knockout. The model 
uses only experimental form factors 
as the input and   
can be readily applied 
to light- and    
medium-mass nuclei.
Pauli blocking effects in these latter nuclei   
are illustrated with the case of the Coulomb interaction.         
Significant effects are noted for   
beam energies below
$\sim$ 350 MeV and for low momentum transfers.   

\vspace{0.5cm} 
$^{\dagger}$\ e-mail address: liu@lanl.gov
 
\vspace{0.25cm}

\pagebreak 
\section{Introduction}{\label {sec.1} }  
The dominant contribution to electron-nucleus reactions at energies  
below the pion production threshold comes from quasielastic electron-nucleus
scattering\cite{Ube}, in which     
a target nucleon is knocked out to the continuum by the incoming electron.   
While exclusive quasielastic  
experiments can provide detailed   
nuclear structure information of the struck nucleon,  
inclusive experiments   
allow us to study various general properties of the 
reaction dynamics~\cite{Cel}-\cite{Kim}. As the final-state interaction 
(FSI) between the knocked-out nucleon and the residual nucleus can affect the calculated 
spectra\cite{Kim}, it must be properly evaluated.   
For exclusive experiments, optical potentials are often   
used to calculate the FSI  
~\cite{Izu}--\cite{Chi}. 
Because these nonhermitian potentials differ from the potential that 
binds the nucleon in the nucleus,  
they generate nucleon scattering wavefunctions
that are not orthogonal to the 
bound-state wavefunction of the nucleon. 
This nonorthogonality leads to    
overestimated (spurious) contribution
to nucleon knockout cross sections 
as the momentum transfer ${\vq}\rightarrow 0$.  
Many methods   
were proposed to restore the orthogonality\cite{Sha}-\cite{Cap}. 
For inclusive experiments, nuclear final states are not measured.   
Hence, in principle, a real-valued potential is to be used for FSI calculations. 
If one solves simultaneously the bound-state and scattering problems 
with a same real-valued potential,  
then the above-mentioned orthogonality difficulty will
not occur. However, very often, particularly in the case of nonrelativistic
treatment of FSI in inclusive experiments, one uses phenomenological 
energy-dependent potentials\cite{Hor}-\cite{Kaw}. These potentials   
differ from the potential that binds the
nucleon. In this respect, the lack of orthogonality exists in practice    
and it is of interest to improve the  
implementation of the required orthogonality in inclusive calculations. 
In this work, we develop a new approach to FSI in inclusive quasielastic 
scattering, which does not need an explicit use of potentials while  
implements the needed orthogonality at all FSI energies on a same footing.   
\pagebreak 
 
Because the distortion of the electron waves  
in the initial and final states can be taken into account by the 
DWBA method and is of no relevance for the discussion 
presented in this work, we will, therefore, use     
plane waves for the electrons so as to show more clearly the effects  
of blocking spurious knockouts in the new approach.  
The theory is developed in Section~\ref{sec.2} and its application is    
given in Section\ \ref{sec.3}.
Discussion  
and conclusions are presented in Section\ \ref{sec.4}.

\section{Electron quasielastic scattering from a nucleus}{\label{sec.2} }  
The one-photon exchange, one-nucleon knockout amplitude,
$ {\bf A}$, is illustrated in 
Fig.1 
where the four-momenta of the on-shell particles (external lines of the
diagram)  
are denoted by  
$p_{_i}=(E_{i},\vp_{i})  $
with $i=(0,1,2,C,A)$. 
The four-momentum of the photon is   
$q = p_{_0} - p_{_2}  \equiv
(\omega, \vq)$. 
With the Bjorken-Drell convention~\cite{Bjo} for the metric, 
single-particle state 
normalization, and reaction cross section,  
the quasielastic scattering differential cross section equals to    
\begin{eqnarray}
\frac{d^2\sigma}{d\Omega_2dE_2} &=& \int 
\frac{(\tpi)^4}{v_{in}} \sum_{spins}
\ \delta^3({\vp}_0+{\vp_A}-{\vp_1}-{\vp_2}-{\vp}_{_C})\delta(E_0+E_A-E_1-E_2-E_{_C}) \nonumber \\  
& &  \ \left(\frac{m_e\Ma}{E_0E_{_A}}\right) 
\ \frac{1}{2(2J_A+1)} \left|{\bf A}\right|^2    
 \frac{|\vp_2|E_2}{(\tpi)^3(E_2/m_e)}\frac{d{\vp}_1}{(\tpi)^3(E_1/{\Mn})}\frac{d{\vp}_{_C}}{(\tpi)^3(E_{_C}/\Mc)}  \ , 
\label{2.1}
\end{eqnarray}  
where   
$v_{in} =  
E_0E_A/\sqrt{(p_0\cdot p_A)^2-p^2_0p^2_A}$   
is the relative velocity in the initial channel, $J_A$ is the spin 
of the target nucleus, and the summation is over   
the spin projections of the external particles.  

As in any Feynman diagram, 
the intermediate particles are off-mass-shell particles. 
This is the case with the intermediate photon,  
the intermediate nucleon, $j$, and the corresponding 
residual nucleus, denoted $\cj$.   
However, it is useful to put the intermediate heavy nucleus, $\cj$,  
on its mass shell and to retain  
only the positive-energy 
spinors of the nucleon $j$.  
This covariant approximation enables one to  
use the bound-state nuclear wavefunctions given by  
traditional nuclear structure theories in which the negative-energy 
component of the wavefunction is not considered.\cite{Liu}  \ \ 
Because the difference among various nuclear masses $M_{_{\cj}}$ is $\ll {\Mn}$, it is 
also useful   
to define $M_{_C}$ as an average of $M_{_{\cj}}$ and substitute  
the former for the latter.     
One thus has     
\begin{figure}[htb]\centering 
%\centerline{\psfig{figure=figuu1.ps,angle=-0.0,height=20cm,width=14cm,clip=}
%}
\mbox{ \epsfysize=130mm 
       \epsffile{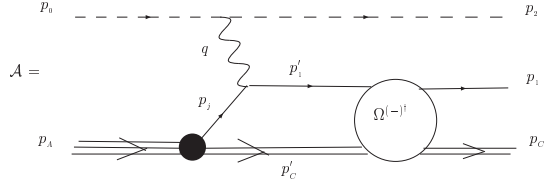} } 
\vspace{-8cm}
\caption{Amplitude ${\bf A}$ for quasielastic scattering. The dashed, 
wavy, solid, and multiple solid lines represent, respectively, the electrons,
the photon, the nucleon, and the nuclei. $\Omega^{(-)^{\dagger}}$ is 
the wave operator for the nucleon final-state interaction. 
A summation over the 
target nucleon label $j$ is understood.}  
\label{Fig1}
\end{figure}  
\vspace{0.25in} 
\begin{eqnarray}
 {\bf A}& =& \left(\frac{ee_pf(q^2)}{q^2}\right)\ 
 \ubar(\vp_2,s_2)\gamma_{\nu}u(\vp_0,s_0)\       
 \sum_{J_j\mu_j} \sum_{J_{_{\cj}}s_{_{\cj}} }  
  \sum_{s_js'_1}
\int \frac{d\vp_j}{(\tpi)^3(E_j/\Mn)(E'_{_C}/\Mc)(E'_1/\Mn) }  
\nonumber \\    
&\times&\ \langle \vp_1\frac{1}{2}s_1;\ \vp_{_C}J_{_C}s_{_C} |\Omega_j^{(-)^\dagger}|\vpp_1\frac{1}{2}s'_1;\ \vpp_{_C}J_{_{\cj}}s_{_{\cj}}\rangle   
\ \langle \vpp_1\frac{1}{2}s'_1|J^{\nu}(0)|\vp_j\frac{1}{2}s_j\rangle\nonumber \\
&\times& \left[
\frac{\langle\vp_j\frac{1}{2}s_j;\ \vpp_{_{\cj}}J_{_{\cj}}s_{_{\cj}}|\Gamma|p_{_A}J_{_A}s_{_A} \rangle}   
{p^0_j-E_j+i\epsilon}\right] \ .   
\label{2.2}
\end{eqnarray} 
In Eq.(\ref{2.2})    
the abbreviated notations $E'_{_C}\equiv E_{_C}(\vpp_{_C}), E_j\equiv E_{N}(\vp_j) $ and 
$E'_1\equiv E_{_N}(\vpp_{_1})$ are used. 
The square of the four-momentum transfer is $q^2=\omega^2 - |\vq|^2$. 
The four-momentum conservation at each 
interaction vertex gives $\vpp_1= \vp_j + \vq$,\  
${p'_1}^0= p^0_j + \omega$ ,\ $\vp_j =\vp_{_A} -\vpp_{_C}$,\ 
and $p^0_j = E_{_A}(\vp_{_A}) - E'_{_C}(\vpp_{_C})$.   
The $J_j,\mu_j$ are the total angular momentum and its third component 
of the $j$-th target proton , and $\sum_{J_j\mu_j} = Z$ being
the total number of the target protons. The $e$ and $e_p$ denote, respectively, 
the electron and proton charges, and the $u(\ubar)$ and $U(\Ubar)$ the 
corresponding spinors. The $f(q^2)$ is the $\gamma pp$ form factor and  
\begin{equation}
\langle \vpp_1\frac{1}{2}s'_1|J^{\nu}(0)|\vp_j\frac{1}{2}s_j\rangle 
= \Ubar(\vpp_1,s'_1)J^{\nu}(q)U(\vp_j,s_j) 
=\int d\vx
\langle \vpp_1\,\frac{1}{2}s'_1|e^{i\vq\cdot\vx}J^{\nu}(\vx)|\vp_j\frac{1}{2}s_j\rangle 
\label{2.2b}
\end{equation}
where $J = (J^0,\vec{J})$ is the electromagnetic current operator. 
For single-nucleon processes one can represent 
the target nucleus as an active nucleon $i$ and a corresponding 
spectator 
residual nucleus $\ci$, {\em i.e.},    
\begin{eqnarray}
 |\vp_{_A}J_{_A}s_{_A}\rangle &= &\sum_{J_iJ_{_{\ci}}}{ F}(J_iJ_{_{\ci}}
; J_{_A})    
 \sum_{s_{_{\ci}}\mu_i }   
{ C}(J_i\mu_i,\ J_{_{\ci}}s_{_{\ci}}|J_{_A}s_{_A})   
|\vp_i;J_i\mu_i\rangle|(\vp_{_A}-\vp_i);J_{_{\ci}}s_{_{\ci}}\rangle\ ;\nonumber \\   
 |J_i\mu_i\rangle &=& 
\sum_{m_i,s_i} { C}(\frac{1}{2}s_i,\ \ell_i m_i|J_i\mu_i)  
 \ \ |\frac{1}{2}s_i\rangle\ |\Phi_{J_i\ell_i m_i}\rangle\ .  
\label{2.4b}
\end{eqnarray}  
Here ${ F}(J_iJ_{_{\ci}};J_{_A}) \equiv [J_i^{\nu-1}(J_{_{\ci}})
J_iJ_{_A}\ |\}\ J^{\nu}_i  J_{_A}]$   
is the coefficient of fractional 
parentage, with $\nu$ being the number of protons in the shell having the 
momentum $J_i$. The ${ C}$'s  
are the Clebsch-Gordan coefficients.
Upon using the bound-state equation 
$G_{0}\Gamma \Phi_{bd} = \Phi_{bd}$ (with  
$G_{0}=(p^0_j-E_j+i\epsilon)^{-1}$ and      
$\Phi_{bd}=|J_j\mu_j\rangle$), one obtains   
the covariant single-particle nuclear wavefunction given by  
\begin{eqnarray} 
\Phi_{\{j\}}(\vec{\lam}_j)      
&\equiv & \frac{\langle\vp_j\frac{1}{2}s_j;\ \vpp_{_{\cj}}J_{_{\cj}}s_{_{\cj}}|\Gamma|\vp_{_A}j_{_A}s_{_A} \rangle}
{p^0_j-E_j+i\epsilon} \nonumber \\    
& = & 
  { F}(J_iJ_{_{\ci}};J_{_A}){ C}(J_j\mu_j,\ J_{_{\cj}}s_{_{\cj}}|J_{_A}s_{_A} ) 
\ { C}(\frac{1}{2}s_j,\ \ell_j m_j|J_j\mu_j)  
\  \Phi_{J_j\ell_j m_j}(\vec{\lam}_j) \ .    
\label{2.4c}
\end{eqnarray}  
Here, $\{j\}$ stands for the ensemble of quantum numbers 
$J_j,\mu_j,J_{_{\cj}},s_{_{\cj}},s_j,\ell_j,m_j$. 
Furthermore, $\vlam_j = \eta \vp_j - {\vpp}_{_{\cj}}/A =
\vp_j - \vp_{_A}/A$  with $\eta =(A-1)/A$ 
is the relative momentum 
between nucleon {j} and the corresponding  
residual nucleus $C(j)$. 

In Eqs.(\ref{2.1})-(\ref{2.4b}), the states $|\ \rangle$ and $\langle\ |$   
are covariantly normalized, namely, 
$\langle\ \vk',s'|\vk,s\ \rangle = (E(\vk)/M)^{1/2}\delta(\vk'-\vk')\delta_{s's}$. 
On the other hand, in  nonrelativistic nuclear theories the states, which we  
denote $|\ \rangle\rangle$ and $\langle\langle\ |,$\  
have the normalization     
$\langle\langle\ \vk',s'|\vk,s\ \rangle\rangle$$=\delta(\vk'-\vk)\delta_{s's}$. 
Hence,   
$|\vk\ \rangle = |\vk\ \rangle\rangle\ (E(\vk)/M)^{1/2}$. 
It follows that $\Phi$ is related to 
its noncovariantly normalized counterpart,  
$\phi$, by   
\begin{equation}
\Phi_{J_j\ell_jm_j}(\vec{\lam}_j)= \left(\frac{E_jE_{\cj}E_A}{{\Mn}\Mc\Ma}\right)^{1/2} 
\phi_{J_j\ell_jm_j}(\vec{\lam}_j) \ ,        
\label{2.5}
\end{equation}
where 
$\phi_{J_j\ell_jm_j}(\vec{\lam}_j) = R_{j_j\ell_j}(|\vlam_j|)Y^{\ell_j}_{m_j}(\hat{\lam_j})$.    
Being dependent on the relative momentum ${\vec{\lam}_j}$, $\phi$ is  
a spectral wave function. Its relation to the corresponding shell-model 
wave function is given in Ref.\cite{Liu2}.   
Upon introducing Eqs.(\ref{2.2})-(\ref{2.4c}) into 
Eq.(\ref{2.1}), one        
can write Eq.(\ref{2.1}) in the following compact form:  
\begin{equation}
\frac{d^2\sigma}{d\Omega_2dE_2} = 
\left(\frac{d\sigma_M}{d\Omega_2}\right) 
\left(\frac{m^2_e}{E_0E_2}\frac{{ L}_{\mu\nu}
{ W}^{\mu\nu} }{ cos^2(\theta_2/2)} \right) \ ,     
\label{2.15} 
\end{equation} 
where 
${ L}_{\mu\nu}  =    
\frac{1}{2}\ \sum_{s_0s_2} 
 \left[\ubar(\vp_0,s_0)\gamma_{\mu}u(\vp_2,s_2) 
\ubar(\vp_2,s_2)\gamma_{\nu}u(\vp_0,s_0)\right]$  
and 
\begin{eqnarray} 
{ W}^{\mu\nu} & = &   
\int 
\frac{(\tpi)^3}{v_{in}}
\ \delta^3({\vp}_0+{\vp_A}-{\vp_1}-{\vp_2}-{\vp}_{_C})\delta(E_0+E_A-E_1-E_2-E_{_C}) 
\ \frac{|\vp_{_2}|\Ma}{E_2  
E_{_A} }|f(q^2)|^2   
 \nonumber \\
& & 
\sum_{\{j\},\{i\}}\int\frac{d\vp_jd\vp_{i}}{(\tpi)^6(E_j/\Mn)(E_i/\Mn)} 
\left(\frac{M_{_C}}{E'_{_{\cj}}}\right)
\left(\frac{M_{_C}}{E''_{_{\ci}}}\right)
\left(\frac{{\Mn}^2}{E_{_N}(\vp_j+\vq)E_{_N}(\vp_i+\vq)}\right)
 \nonumber \\ 
& &\frac{1}{2} \sum_{s'_1s''_1} \langle \vp_{i}\frac{1}{2}s_i\ |J^{\mu}(0)|\ (\vp_i+\vq)\frac{1}{2}s''_1\rangle  
\langle\ (\vp_j+\vq)\frac{1}{2}s'_1\ |J^{\nu}(0)|\ \vp_{j}\frac{1}{2}s_j\rangle  
\ \Phi^{*}_{\{i\}}(\vec{\lam}_i)  
 \Phi_{\{j\}}(\vec{\lam}_j)   
           \nonumber \\         
& &\langle(\vp_i+\vq)\frac{1}{2}s''_1;\ \vp\,\,''\!\!\!\!_{_C}J_{_{\ci}}s{_{\ci}}|\Omega_i^{(-)} 
 { I}(1,C){\Omega_j^{(-)}}^{\dagger}    
 |(\vp_j+\vq)\frac{1}{2}s'_1;\ \vpp_{_C}J_{_{\cj}}s_{_{\cj}}\rangle \ .  
\label{2.24}
\end{eqnarray}
The  
$d\sigma_{_M}/d\Omega_2$ is the Mott differential cross section and is given by  
\begin{equation}
\frac{d\sigma_{_M}}{d\Omega_2} = 
 \frac{e^2e_p^2}{(\tpi)^2}\frac{E^2_2}{(q^2)^2}cos^2(\theta_2/2) \ .  
\label{2.17}
\end{equation}
In Eq.(\ref{2.24})    
\begin{equation} 
{ I}(1,C)\equiv \sum_{s_1J_{_C}s_{_C}}\ \int    
|\vp_1\frac{1}{2}s_1;\vp_{_C}J_{_C}s_{_C}\rangle  
 \langle\vp_1\frac{1}{2}s_1; \vp_{_C}J_{_C}s_{_C}|   
\ \frac{d{\vp}_1d{\vp}_{_C}}{(\tpi)^6(E_1/{\Mn})(E_{_C}/M_{_C})}= 1\ ,  
\label{2.7} 
\end{equation} 
as a result of the completeness of free two-particle states. Consequently,   
\begin{equation}
 \Omega_i^{(-)}{ I}(1,C)\Omega_j^{(-)^\dagger} =  
\Omega_i^{(-)}\Omega_j^{(-)^\dagger}\delta_{ij} .    
\label{2.8a} 
\end{equation}
The appearance of $\delta_{ij}$ is a consequence of  
one-step reaction process in which the residual nucleus acts as a spectator.
Because the nucleon $j$ and the 
residual nucleus can form bound states,    
the unitary equation  of the wave operators is\cite{Rod}     
\begin{equation}  
 \Omega_i^{(-)}\Omega_j^{(-)^\dagger}\delta_{ij} =  \left( \  
  {\bf 1} - \Gamma_j \right)\delta_{ij} \ ,  
\label{2.8}
\end{equation} 
with 
\begin{equation} 
 \Gamma_j =  \sum_{n=0}^{n_{max}} |n_{\{j\}}\rangle\langle n_{\{j\}}| 
\equiv \sum_{n} \Gamma^{(n)}_j  \ .     
\label{2.8b} 
\end{equation} 
Here, $\Gamma_j^{(n)}$ denotes the projector to the bound state 
$|n_{\{j\}}\rangle$, with $n=0$ denoting    
the nuclear ground state and $n\neq 0$ the  
nucleon-emission-stable (NES) excited nuclear states.    
In the single-step reaction model, $|n_{\{j\}}\rangle = 
 |J^{(n)}_j\rangle\otimes|J_{_{C(j)}}\rangle $. Here, a nucleon $j$ is 
lifted from its ground-state orbital (denoted $J_j$) to an excited orbital 
( denoted $J_j^{(n)}, n\neq 0$).  

The projectors $\Gamma^{(n)}_j$ have the properties  
${\Gamma}^{(n)}_j = {{\Gamma}^{(n)}_j}^{\dagger}$ and 
${\Gamma}^{(n)}_j {{\Gamma}^{(m)}_j}^{\dagger} = \Gamma^{(n)}_j \delta_{nm}$.
These  properties allow us to rewrite Eqs.(\ref{2.8}) and (\ref{2.8b}) 
as  
\begin{equation}
 \Omega_i^{(-)}\Omega_j^{(-)^\dagger}\delta_{ij} =    
  \left( \  
  {\bf 1} - \Gamma_j \right) \delta_{ij} =  
 \left( {\bf 1} - \sum_{n=0}^{n_{max}} |n_{\{j\}}\rangle\langle n_{\{j\}}| 
\ {\bf 1}\ |n_{\{j\}}\rangle\langle n_{\{j\}}|\ \right)\delta_{ij} \ .  
\label{2.9}
\end{equation} 
This last equation defines the doorway model of the final-state nucleon-nucleus 
interaction. 

Using Eqs.(\ref{2.8a})-(\ref{2.9}) for the last line of Eq.(\ref{2.24}),  
one obtains,    
after some 
angular-momentum recoupling algebra, that    
\begin{eqnarray} 
{ W}^{\mu\nu}  =   
\int \frac{d\vp_1d\vp_{_C}}{(\tpi)^3v_{in}}   
&\ & \delta^3({\vp}_0+{\vp_A}-{\vp_1}-{\vp_2}-{\vp}_{_C})\delta(E_0+E_A-E_1-E_2-E_{_C}) 
    \frac{|\vp_2|}{E_2}|f(q^2)|^2 \nonumber \\   
&\times & \left(\ \Xi^{\mu\nu}_I\ -\ \Xi_{II}^{\mu\nu}\right) \ ,   
\label{2.12} 
\end{eqnarray} 
with
\begin{equation} 
\Xi_{I}^{\mu\nu} = \frac{1}{2}\sum_{s_1}\sum_{J_j\mu_j\ell_jm_js_j} 
\ \langle\langle\ \vp_{j}\frac{1}{2}s_j\ |J^{\mu}(0)| \vp_1\frac{1}{2}s_1\rangle\rangle  
\langle\langle\ \vp_1\frac{1}{2}s_1\ |J^{\nu}(0)|\ \vp_{j}\frac{1}{2}s_j\rangle\rangle     
\ \ |\phi_{\{j\}}(\vec{\lam}_j)|^2  \ ,        
\label{2.13}
\end{equation} 
and 
\begin{eqnarray}
\label{2.14} 
\Xi_{_{II}}^{\mu\nu} & = &  
 \sum_{n=0}^{n_{max}}\frac{1}{2}\sum_{s's''}\sum_{J_j\mu_j\ell_jm_js_j}\  
|\phi_{\{j\}}^{(n)}(\vlam)|^2    
     \nonumber \\  
&\times & \left[ \int \frac{d\vp_j}{(\tpi)^{3}} 
\ \phi^{*}_{\{j\}}(\vec{\lam}_j)   
\ \langle\langle\ \vp_j\frac{1}{2}s_j\ |J^{\mu}(0)|\ (\vp_{j}+\vq)\frac{1}{2}
s''\ \rangle\rangle 
\ \phi^{(n)}_{\{j\}}(\vec{\lam}_j+\eta\vec{q})\right]   
      \nonumber \\ 
&\times &\left[ \int \frac{d\vp_i}{(\tpi)^3} 
\ \phi^{(n)*}_{\{j\}}(\vec{\lam}_i+\eta\vec{q})        
\ \langle\langle\ (\vp_i+\vq)\frac{1}{2}s'\ |J^{\nu}(0)|\ \vp_{i}\frac{1}{2}s_j\ \rangle\rangle 
\ \phi_{\{j\}}(\vec{\lam}_i)\right] \ ,  
\end{eqnarray} 
where $\vlam = \eta\vp_1 -A^{-1}\vp_{_C}$  
is the relative momentum of the nucleon-residual nucleus
system in the final state. The momentum conservation at the 
$\gamma pp$ vertex gives  
$\vlam =\vlam_j + \eta\vq$.  
For succinctness of notation, Eqs.(\ref{2.13}) and (\ref{2.14})  
are expressed in terms of noncovariantly normalized  
nuclear wave functions $\phi_{\{j\}}$, and 
noncovariant states $\langle\langle\ |$ and  
$|\ \rangle\rangle$. Consequently,
various normalization factors, of the form $(E/M)$, are implicit.  

\begin{figure}[htb]\centering 
%\centerline{\psfig{figure=figuu2.ps,angle=-0.0,height=20cm,width=14cm,clip=}
%}
\mbox{ \epsfysize=120mm
        \epsffile{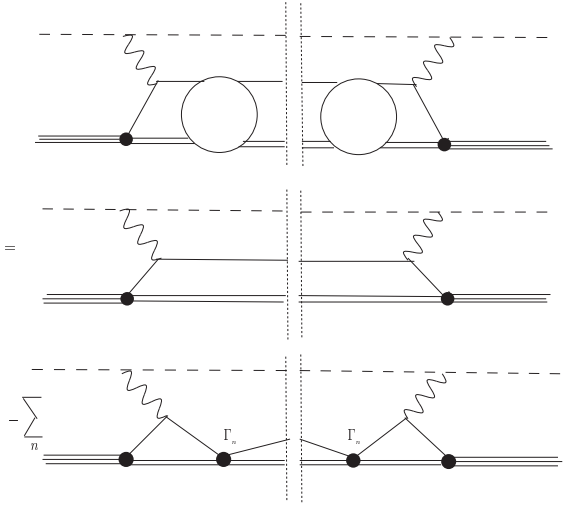} } 
\vspace{-4cm}
\caption{The doorway model for Pauli-blocking corrections.  
As in Fig.\ref{Fig1}, the subscript $j$ and a summation over it on both side of
the graphic equation is implied.}  
\label{Fig1a}
\end{figure}

Eq.(\ref{2.12})
is illustrated   
in Fig.\ref{Fig1a}. Its physics content is as follows.   
The $\Xi_{_I}$  
leads to cross sections 
obtained with using plane waves in the final state.    
The $\Xi_{_{II}}$ gives the cross sections for the struck nucleon to  
remain bound. The subtraction of $\Xi_{II}$ from $\Xi_I$ 
corrects the spurious contribution arising from using plane waves.   
As we shall see, at $\vq$=0 the subtraction is total; in other words,   
the spurious proton knockout is completely blocked. 
\pagebreak
\vspace{-26cm}  
Using the well-known Lorentz-invariant parametrization\cite{Dre},\cite{Wal} of 
the response tensor ${ W}^{\mu\nu}$,    
one obtains        
\begin{equation} 
\frac{d^2\sigma}{d\Omega_2dE_2} = \frac{d\sigma_{_M}}{d\Omega_2}
\left[ \left(\frac{(q^2)^2}{|\vq|^4}\right)R_L(\omega,|\vq|)
 + \left(\frac{q^2}{2|\vq|^2} - tan^2(\theta_2/2)\right)R_T(\omega,|\vq|)\right] 
\label{2.21} 
\end{equation}
in the laboratory frame. Here, $R_T$ and $R_L$ are, respectively, 
the transverse and longitudinal response functions with    
$ R_T =  \sum_{_{\lam=\pm 1}}    
({\vec{e}\ }^{\dagger}_{\vq,\lam})_{_i} { W}^{ij}
\left({\vec{e} }_{\vq,\lam}\right)_j$\ $(i,j=1,2,3)$  
and   
$R_L  = { W}^{00} = \Xi_I^{00} - \Xi_{II}^{00}$.   

\vspace{-0.2in} 
\section{Effects of Pauli Blocking }{\label{sec.3} }    
To illustrate the blocking of spurious nucleon knockout in the 
doorway model,    
let us consider the Coulomb scattering only. 
In this latter case, $J^{\mu}= 
(\hat{\rho}, \vec{0})$. 
Hence, $R_T$=0 and  
\begin{equation}
\frac{d^2\sigma}{d\Omega_2dE_2} = \frac{d\sigma_{_M}}{d\Omega_2}
\frac{(q^2)^2}{|\vq|^4} R_L \ .  
\label{2.27b} 
\end{equation}
In the second quantization  
$\hat{\rho}(\vx) = \hat{\psi}^{\dagger}(\vx)\hat{\psi}(\vx).$    
Upon using the nonrelativistic two-component
proton field  
$\hat{\psi}(\vx) = (\tpi)^{-3/2}\int d{\vk} 
\ \sum_{\xi} 
e^{i\vk\cdot\vx}\ a_{\vk,\xi} \chi_{\xi}, $    
one finds that 
the two matrix elements of $J^0$ in the 
first square brackets in $\Xi_{I}^{00}$ equal 
to $\delta_{s_1s_j}$ while the two matrix elements 
of $J^0$ in $\Xi_{II}^{00}$ become, respectively,
 $\delta_{s''s_j}$ and 
$\delta_{s's_j}$. Consequently,   
\begin{equation}
R_L = \int \frac{d\vK}    
{v_{in}}
\ \delta^3(\vq +\vp_{_A}-\vK) 
\frac{|\vp_2|}{E_2}|f(q^2)|^2\ R(\omega,|\vq|) 
\label{2.31}
\end{equation}
with  
\begin{eqnarray}
& &\hspace{1.5in}  R(\omega,|\vq|) =    
\int \frac{d\vlam}{(\tpi)^3} 
\delta(\omega + E_{_A}- E_{1} - E_{_C}) \nonumber \\    
& & \left[\sum_j \left(  |\phi_{\{j\}}(\vec{\lam_j})|^2 -
 |\phi_{\{j\}}(\vlam)|^2\ |F_{\braj}^{00}(\vq)|^2\right)  
-{\sum_j}\ ' \sum_{n\neq 0} |\phi_{\{j\}}^{(n)}(\vlam)|^2\ |F_{\braj}^{0n}(\vq)|^2\right]\ .  
\label{2.32}
\end{eqnarray}
In obtaining Eqs.(\ref{2.31}) and (\ref{2.32})    
we used the relations $d\vp_1d\vp_{_C}=d\vK d\vlam$ 
(with $\vK\equiv \vp_1+\vp_{_C}$) and 
\begin{equation}
 \int \frac{d\vp_i}{(\tpi)^3} 
\phi^{(n)*}_{\{j\}}(\vec{\lam}_i+\eta\vq)   
\ \phi_{\{j\}}(\vec{\lam}_i) = \int d\vec{r}_i\ e^{i\vq\cdot\vec{r}_i}\ \psi^{(n)*}_{\{j\}}(\vec{r}_i)\ \psi_{\{j\}}(\vec{r}_i) =      
F^{0n}_{\braj}(\vq) \ .  
\label{2.32b}
\end{equation} 
The $\sum\ '$ in Eq.(\ref{2.32}) indicates that not 
every target proton is involved
in a $0\rightarrow n$ transition. Hence, 
${\sum_j}'\ 1\equiv Z' \leq Z$. The $\delta$ function 
in Eq.(\ref{2.32}) constrains the energy loss $\omega$ 
and makes $\omega$ depend on $\vec{\lam}^2$ and 
the average proton separation energy $B = M_1 + M_{_C} - M_{_A}$.
   
In Eq.(\ref{2.32}), $F_{\braj}^{00}(\vq)\equiv F_{\braj}^{g.s.\rightarrow g.s.}(\vq)\ $
is the nuclear (ground-state) form factor of the $j$-th 
proton with the property $F_{\braj}^{00}(0)=1$. 
For $n\neq 0$, $F_{\braj}^{0n}(\vq)\equiv F_{\braj}^{g.s.\rightarrow n}(\vq)$
are the transition form factors, and   
$F_{\braj}^{0n}(0)=0.$   
Consequently,
when $\vq\rightarrow 0$, $R \rightarrow 0$; {\it i.e.,}    
the knockout of a target proton is completely blocked at $\vq=0$.  
We have noted that 
experimental form factors are not parametrized with respect to 
an individual proton but rather with respect to the      
whole nucleus as a function of $|\vq|$. (Henceforth,  
$|\vq|$ is denoted as $q$ for a succinct notation.)  
It is, therefore, appropriate to introduce     
\begin{eqnarray} 
F_{\braj}^{00}(\vq)& =& \frac{1}{Z}F_A^{00}(q)\equiv F^{00}(q)\ , \nonumber \\     
F_{\braj}^{0n}(\vq)& =& \frac{1}{Z'}F_A^{0n}(q) \equiv F^{0n}(q)\ \ (n\neq 0) \ .   
\label{2.33}
\end{eqnarray}
The $q$-dependence of PBC can be obtained by integrating
over all energy loss in Eq.(\ref{2.32}).
Using the completeness  
relation  
\begin{equation}
\int \frac{d\vlam}{(\tpi)^3} |\phi_{\braj}(\vlam_j)|^2 =  
\int \frac{d\vlam_j}{(\tpi)^3} |\phi_{\braj}(\vlam_j)|^2 = 1 \ ,   
\label{2.33b} 
\end{equation} 
one obtains     
\begin{equation}
\label{2.34}
  \int d\omega\ R(\omega,q)  
 = Z\left( 1 - |F^{00}(q)|^2 - \beta\sum_{n\neq 0}^{n_{max}}
|F^{0n}(q)|^2\right)  
\equiv ZL(q)\ .  
\end{equation}
The ratio $\beta\equiv Z'/Z$ depends on nuclear excitation mechanisms.  
The function $L(q)$   
gives the probability for a struck proton to leave  
the nucleus. Eq.(\ref{2.34}) shows how the 
doorway and Fermi gas models differ.  In the Fermi gas model, the nucleon 
density distribution $|\psi(\vp_j)|^2$, is assumed 
to be $\theta(|\vp_j|-k_F)$ where $k_F$ is the Fermi momentum. 
Because of the Pauli principle, this box-type momentum-space density 
distribution 
blocks $\psi(\vp_j)\rightarrow 
\psi(\vp_j+\vq)$ transitions  
whenever $|\vp_j+\vq|\leq k_F$.      
For realistic density distributions, there is no such sharp momentum cutoff  
in transitions. Instead,     
the $\psi(\vp_j)$ to $\psi^{(n)}(\vp_j+\vq)$ transition can occur
at any given $\vq$ with the probability  
$|F^{0n}(q)|^2$. 
Hence,    
$|F^{00}(q)|^2 + \beta\sum_{n\neq 0}  
\ |F^{0n}(q)|^2 $  
is the probability that the struck nucleon remains bound. With a minus
sign in front of this last quantity,    
the second and third terms in Eq.(\ref{2.34}) give    
the blocking correction to nucleon knockout in a realistic nucleus. We
name this  
correction the Pauli-blocking correction (PBC) because it is 
a consequence of the Pauli exclusion principle.    
 
A comment on Eq.(\ref{2.34}) is in order. 
While form factors $F^{00}$  
have been determined experimentally for a large number of nuclei,  
experimental information on transition form factors $F^{0n}\ (n\neq0)$ 
is much less systematic.    
However, in nuclei with mass number $A\leq 5$ there is no  
NES excited states. Consequently, only the term $| 
F^{00}|^2$  
is needed in Eq.(\ref{2.34}). The $L(q)$ can, therefore, be calculated 
exactly for these light nuclei with the use of   
experimental form factors. 

\begin{figure}[htb]\centering 
\mbox{   \epsfysize=12cm  
        \epsffile{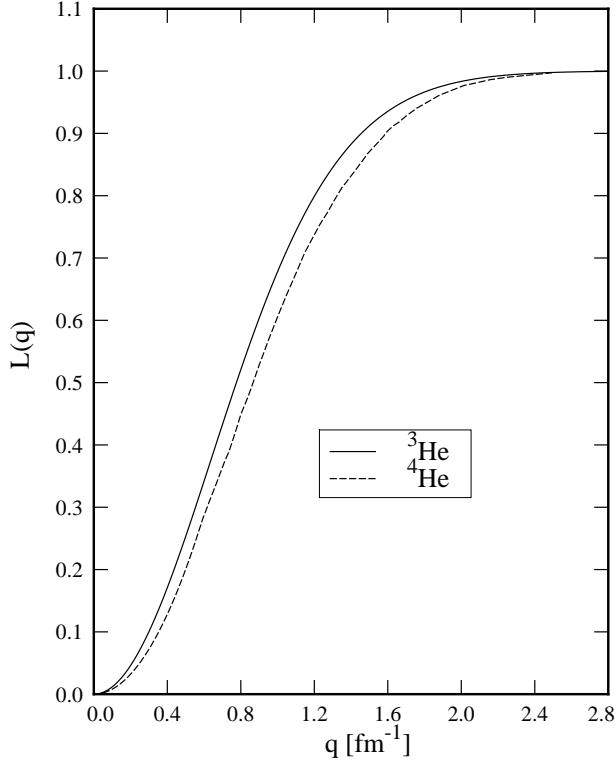}
}
\caption{Function 
$L(q)=1 - |F^{00}(q)|^2$ for  
nuclei $^3$He and $^{4}$He.} 
\label{Fig3}
\end{figure}

\begin{figure}[htb]\centering 
\mbox{   \epsfysize=12cm  
        \epsffile{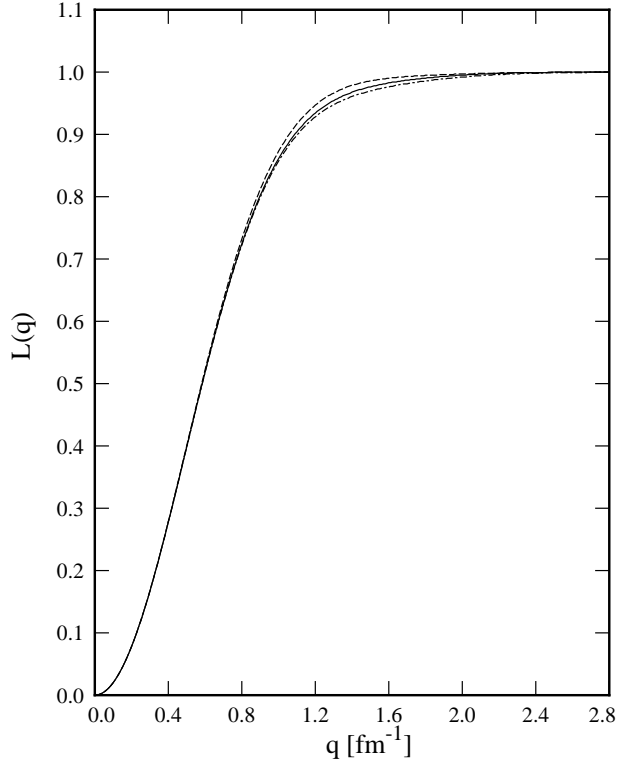}
}
\caption{Functions $L(q)$ for $^{12}C$. Dashed curve: $L=1-|F^{00}|^2$. 
Solid curve: $L= 1 - |F^{00}|^2 - |F^{0,2+}|^2$. Dot-dashed curve:
$L= 1 - |F^{00}|^2 - \sum_n |F^{0n}|^2$   
($ n= 2^+, 1^-, 3^-, 4^+$).}     
\label{Fig4}
\end{figure}

In Fig.\ref{Fig3},\  the functions    
$L(q)=1-|F^{00}(q)|^2$   
for two light nuclei are shown.  
In both cases 
$L(q)=0$ at $q=0$ and $L(q)\rightarrow 1$ when   
$q > 2.7 $fm$^{-1}$. 
Graphically, the PBC is represented   
by $1-L(q)$ 
which is the vertical distance between the curve and the 
horizontal line passing through $L(q)$=1.  Fig.3 shows the PBC 
is complete ({\it i.e.,} 100\%) at q=0 and how 
it decreases with increasing $q$. .     
Since there is only one bound state in $^3$He and 
$^4$He (the ground states),  
$1-|F^{00}(q)|^2$ represents an exact calculation 
of $L(q)$ for these nuclei.  
%\vspace{-1cm} 
In nuclei with mass number $A \geq 6$, there are NES states and its 
number increases with $A$.
To illustrate the effects of NES states in $1p$-shell nuclei, 
we show in Fig.\ref{Fig4} the function $L(q)$ of $^{12}$C,    
assuming $\beta=1$ in Eq.(\ref{2.34}).    
The PBC effects due to $|F^{00}|^2$ and $(\ |F^{00}|^2 
$ + $|F^{0,2^+}|^2\ )$ are given, respectively, 
by the dashed and solid curves 
in the figure.   
Here $2^+$ is the 4.44 MeV ($T=0$) excited state. 
The dot-dashed curve further includes the PBC arising from   
transitions to the NES states~\cite{Fla}--\cite{Ahm}
at  
7.12 MeV ($1^-, T=0$), 9.64 MeV ($3^-,T$=0), 
and 14.1 MeV ($4^+, T$=0). 
Since the proton separation energy in $^{12}$C is 
15.11 MeV, the inclusion of these four states should take into
account most of the NES transition strength. As one can see from Fig.\ref{Fig4}, 
the most
important effects of $|F^{0n}|^2 (n\neq 0)$ 
comes from the transition
to the first $2^+$ excited state at 4.44 MeV. The inclusion of other three states 
brings in only small additional effects. One could expect  
that, in general, only a limited number of transitions to NES states needs to be 
considered in medium-mass nuclei. 
%\vspace{-1cm} 
The relative importance of PBC effects due to different doorway channels can 
be evaluated from comparing the corresponding $\int L(q)dq$. 
We have found that $\int (1-|F^{00}(q)|^2)dq$ (integration of the dashed 
curve)   
differs from  $\int (1-|F^{00}(q)|^2-\sum_n|F^{0n}(q)|^2)dq,$ $(n=2^+,1^-,
3^-,4^+)$ (integration of the dot-dashed curve) by less than 
2\%. In the following calculations of PBC in $^{12}$C, 
we will, therefore, use 
the term $|F^{00}|^2$ only. 
   
\begin{figure}[htb]\centering 
\mbox{   \epsfysize=12cm  
        \epsffile{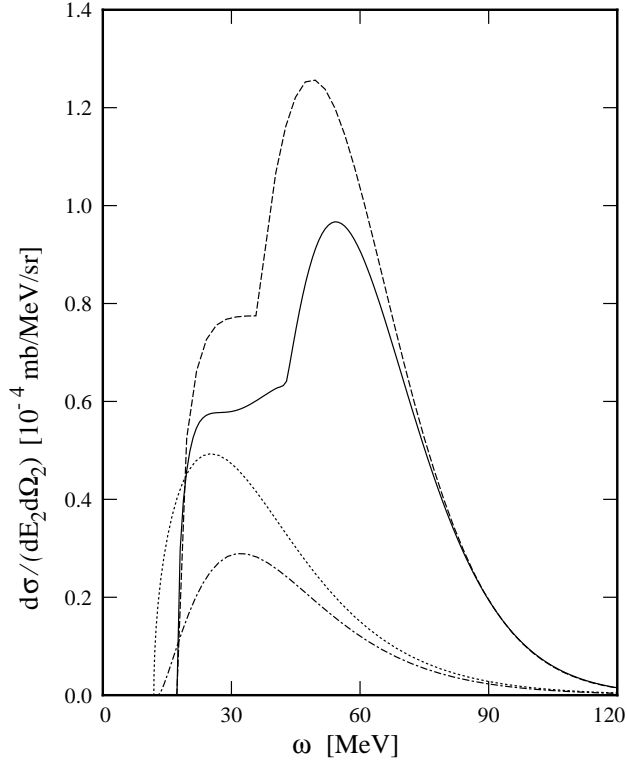}
}
\caption{Inclusive lab. cross sections 
$d\sigma/(dE_2d\Omega_2)$  
at $E_0=200$ MeV 
and $\theta_2=60^o$  
as a function of energy loss $\omega$. Dotted curve:  
$^3$He without PBC.\ Dot-dashed curve: $^3$He with PBC.\ Dashed cureve: $^{12}$C 
without PBC.\ Solid curve: $^{12}$C with PBC. }   
\label{Fig5}
\end{figure}

In Fig.\ref{Fig5}, we show PBC effects on   
inclusive cross sections of quasielastic scattering  
from $^3$He and $^{12}$C  
at $E_0=200$ MeV and $\theta_2= 60^o$ as a function of 
the energy loss $\omega$. For $^{12}$C, realistic separation energies $B_p=15$ and 
$B_s=35$ MeV were used respectively for the $1p-$ and $1s-$shell protons.
These shell-dependent separation energies give rise to  
the shoulder in the $^{12}$C spectra. As we can see,  
the PBC is significant in both nuclei. 
%\vspace{-1cm} 
To quantify the integrated PBC effects on the cross section, let us define 
\begin{equation} 
\delta = \frac{\left(d\sigma/d\Omega_2\right)^{no PBC} -     
          \left(d\sigma/d\Omega_2\right)^{PBC} }       
         { \left(d\sigma/d\Omega_2\right)^{no PBC} }  \ .      
\label{2.39} 
\end{equation} 
The values of $\delta$ in $^3$He and $^{12}$C are given in Table~\ref{tabI}  
where a blank entry represents a $\delta < $1~\%.   
\begin{table}
\begin{center}
\caption{ Pauli blocking correction $\delta$ [\%].  }
\label{tabI} 
%\begin{center} 
\begin{tabular}  
 {|c|| c|c|c || c|c|c|}\hline     
E$_0$ [MeV] & $^3$He:\ \ \ $\theta_2=30^o$ & $\theta_2=45^o$ & $\theta_2=60^o$ 
          & $^{12}$C:\ \ \ $\theta_2=30^o$ & $\theta_2=45^o$ & $\theta_2=60^o$ \\ \hline  
200 &\ \ \ \   78 & 58 & 40 & \ \ \ \   63 & 37 & 18          \\ \hline 
350 &\ \ \ \   44 & 18 &  6 & \ \ \ \   21 & 3 &          \\ \hline
500 &\ \ \ \   18 &  3 &    & \ \ \ \    3 &   &          \\ \hline   
\end{tabular}
\end{center}
\end{table}
As one can see from the table, the PBC decreases 
with increasing energy and scattering angle and becomes negligible at  
$E_0=500$ MeV and $\theta_2=60^o$. Indeed, we have noted that 
under this latter experimental condition
the momentum transfers $|\vq|$ contributing to the bulk of the 
cross sections are greater than  
2 and 2.4 fm$^{-1}$ in $^3$He and $^{12}$C, respectively. 
These large $q$ lead to  
negligible PBC     
 (see Figs.\ref{Fig3} and \ref{Fig4}).    

\begin{figure}[htb]\centering 
\mbox{   \epsfysize=12cm  
        \epsffile{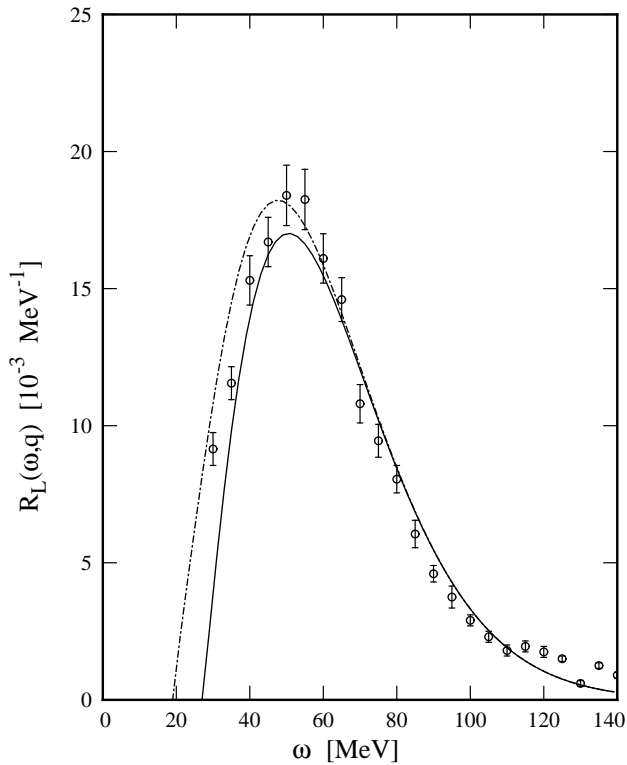}
}
\caption{  
Longitudinal response functions of $^3$He at $|\vq|$=300 MeV/c.
Dot-dashed curve: without PBC. Solid curve: with PBC. Data are from 
Ref.29.}  
\label{Fig6}
\end{figure}

The $R_L$ of $^3$He and $^{12}$C have been measured at $|\vq|=$ 
300 MeV/c\cite{Dow},\cite{Bar}. Since at q= 300 MeV/c the effect of PBC in 
$^{12}$C is unimportant (see Fig.\ref{Fig4}), 
we compare, therefore, in Fig.\ref{Fig6} 
the longitudinal response functions  
of $^3$He given by the doorway model 
at $|\vq|$=300 MeV/c with  
the data\cite{Dow}. 
As one can see, the PBC is 
very important at small $\omega$'s. The inclusion of PBC  
improves the position of the peak of 
the calculated spectrum. 
\vspace{-1cm} 
  
\section{Discussion and Conclusions}{\label{sec.4}  }  
By using the unitarity equation of the wave operators,  
we have developed a doorway model for the nucleon final-state 
interaction in inclusive quasielastic scattering. 
The model does not rely on the use of potentials;
the inputs to the 
calculation are the
experimentally determined  
form factors. 
For nuclei with mass numbers $A\leq 5$, 
the doorway calculation is exact. For $1p-$shell nuclei such as 
$^{12}$C the model can be calculated to a very good approximation with only using 
the measured ground-state (g.s.) nuclear form factor. 
One could expect that this latter approximation equally holds for 
$1d-$ and $1f-$shell nuelci. 
At the present time the application of the doorway model to heavy-mass
nuclei is hindered by the lack of a systematic experimental knowledge of the 
NES transition form factors in these nuclei. It is worth finding out whether 
the use of a few important experimentally known NES form factors 
would suffice.  
Further studies are called for. Our study shows that the Pauli
blocking of spurious nucleon knockout is
important when the electron energies are below 350 MeV (Table~\ref{tabI}). 
The PBC is   
also important when momentum transfers are small.   
The doorway approach derived in this work     
represents a useful alternate to the various FSI  
approaches proposed in the literature. It is calculationally simple 
and can be easily applied to 
the study of inclusive quasielastic scattering from light and medium-mass 
nuclei.   

\vspace{1.0cm} 
%{\bf REFERENCES} 

\end{document}